%% file: SigDLA.tex
\documentclass[conference]{IEEEtran}
\IEEEoverridecommandlockouts
\usepackage{cite}
\usepackage{hyperref}
\usepackage{booktabs} 
\usepackage{tabularx}
\usepackage{ragged2e}
\usepackage{amsmath,amssymb,amsfonts}
\usepackage[ruled,norelsize,vlined,linesnumbered]{algorithm2e}
\usepackage{multirow}
\usepackage{graphicx}
\usepackage{textcomp}
\usepackage{xcolor}

\usepackage{url}
\usepackage{cite}
\usepackage{hyperref}

\usepackage{array}
\usepackage{textcomp}
\usepackage{stfloats}
\usepackage{verbatim}
\usepackage{subfigure}

\begin{document}
	
\title{SigDLA: A Deep Learning Accelerator Extension for Signal Processing\\
}

\author{
\IEEEauthorblockN{Fangfa Fu$^{1}$, Wenyu Zhang$^{1,2}$, Zesong Jiang$^{2,4}$, Zhiyu Zhu$^{1}$, Guoyu Li$^{2,3}$, Bing Yang$^{5}$, \\ Cheng Liu$^{2,3}$\IEEEauthorrefmark{1}, Liyi Xiao$^{1}$, Jinxiang Wang$^{1}$, Huawei Li$^{2,3}$, Xiaowei Li$^{2,3}$}
	\IEEEauthorblockA{
		$^{1}$Department of Microelectronics Science and Technology, Harbin Institute of Technology, Harbin, China
	}
	\IEEEauthorblockA{
		$^{2}$SKLP, Institute of Computing Technology, Chinese Academy of Sciences, Beijing, China
}
	\IEEEauthorblockA{
	$^{3}$Dept. of Computer Science, University of Chinese Academy of Sciences, Beijing, China
 }
    \IEEEauthorblockA{
		$^{4}$Institute of Advanced Technology, University of Science and Technology of China, Hefei, China
	}
	\IEEEauthorblockA{
		$^{5}$Dept. of Computer Science and Technology, Harbin University of Science of Technology, Harbin, China
	}
	\IEEEauthorblockA{liucheng@ict.ac.cn, 22s121114@stu.hit.edu.cn, jiangzesong@mail.ustc.edu.cn}
	\vspace{-2.8em}
\thanks{*Cheng Liu is the corresponding author.}
 \thanks{This work is supported by the National Key R\&D Program of China under Grant (2022YFB4500405), and the National Natural Science Foundation of China under Grant 62174162.}
}

\maketitle

\begin{abstract}
	Deep learning and signal processing are closely correlated in many IoT scenarios such as anomaly detection to empower intelligence of things. Many IoT processors utilize digital signal processors (DSPs) for signal processing and build deep learning frameworks on this basis. While deep learning is usually much more computing-intensive than signal processing, the computing efficiency of deep learning on DSPs is limited due to the lack of native hardware support. In this case, we present a contrary strategy and propose to enable signal processing on top of a classical deep learning accelerator (DLA). With the observation that irregular data patterns such as butterfly operations in FFT are the major barrier that hinders the deployment of signal processing on DLAs, we propose a programmable data shuffling fabric and have it inserted between the input buffer and computing array of DLAs such that the irregular data is reorganized and the processing is converted to be regular. With the online data shuffling, the proposed architecture, SigDLA, can adapt to various signal processing tasks without affecting the deep learning processing. Moreover, we build a reconfigurable computing array to suit the various data width requirements of both signal processing and deep learning. According to our experiments, SigDLA achieves an average performance speedup of 4.4$\times$, 1.4$\times$, and 1.52$\times$, and average energy reduction of 4.82$\times$, 3.27$\times$, and 2.15$\times$ compared to an embedded ARM processor with customized DSP instructions, a DSP processor, and an independent DSP-DLA architecture respectively with 17\% more chip area over the original DLAs.
\end{abstract}


\IEEEpeerreviewmaketitle
\input{01_intro}
\input{02_Relatedwork}
\input{03_SigDLA_Architecture}

\input{04_Computing_Array}

\input{05_Data_Shuffling}
\input{06_Evaluation}
\input{07_Conclusion}

\bibliographystyle{IEEEtran}
\bibliography{SigDLA}

\end{document}

%% file: 01_intro.tex
\section{Introduction}
Deep learning has been demonstrated to be successful in numerous domains of applications, is increasingly adopted in IoT devices to enable intelligence of things under various scenarios, such as anomaly detection and status monitoring \cite{li2024digital} \cite{xu2023lamb} \cite{rui2024signal}. While many of these IoT devices rely on sensors to capture physical signals such as vibration and temperature for the detection or monitoring, signal processing that focuses on denoising and transformation is usually applied with the deep learning processing for more effective inference \cite{li2024digital} \cite{xu2023lamb} \cite{rui2024signal} \cite{intr2}. Although specific signal processing algorithms and deep learning models may vary across different IoT applications, they are generally required at the same time and involve massive data transfer between them due to consecutive processing.

However, many IoT computing engines utilize a DSP processors to perform signal processing and build deep learning systems on top of the DSP processor \cite{DSPwithDL1}  \cite{DSPwithDL3} \cite{DSPwithDL5} \cite{zhang2022comprehensive} or even a general-purpose processor (GPP)\cite{david2021tensorflow} \cite{cmix} \cite{mcunet} \cite{liu2024tinyts}, which fails to achieve energy-efficient deep learning due to the lack of native hardware support. Particularly, deep learning is usually more compute-intensive and memory-intensive compared to signal processing. Hence, implementing deep learning on DSP is suboptimal for IoT devices that feature both signal processing and deep learning. Some of the recent IoT processors\cite{CPU_DSP_DL} also have custom deep learning processors embedded and seated along with DSP processors. Essentially, they have deep learning and signal processing performed on independent DLAs and DSP, respectively, for the sake of optimized energy efficiency. However, the intelligent signal analysis demands non-trivial data transfer between the DSP processor and deep learning processor, which will incur substantial communication overhead in terms of power and latency. Moreover, independent accelerators with private on-chip buffers and computing arrays inevitably consume larger chip area and lead to higher chip price\cite{AIOT_XIAOLV}, which is usually unacceptable for cost-sensitive IoT devices. 

In this paper, we propose a novel approach to extend signal processing on top of a typical DLA and build a unified accelerator called SigDLA. We note that both DLA and DSP utilize MAC arrays for computation, and we aim to map two different workloads onto the same MAC array. The major barriers that hinder the mapping of signal processing on deep learning computing array are roughly the shuffled processing like butterfly operations in FFT and the larger data width which usually depends on the sensor resolution. For shuffled operations \cite{fft_shuffle} \cite{shuffling1} \cite{shuffling2} \cite{shuffling3} \cite{shuffling4}, we propose a data shuffling fabric and have it inserted between on-chip memory and the DLA computing array. The shuffling fabric reorganizes the shuffled operations such that they can be converted to standard tensor operations to fit the regular computing array in DLAs. The shuffling fabric is programmable to suit different data reorganization requirements of various irregular operations in signal processing. To handle the wide data width, we build a serial processing element-based MAC array to support tensor operations with variable data width, which has been intensively explored in prior works\cite{BitBlade}\cite{BitFusion}. 

The proposed architecture can also be utilized in compute-intensive tasks beyond signal processing. The benefits of the unified architecture are multi-folded. Firstly, it achieves optimized performance of deep learning which usually dominates the execution time of intelligent IoTs and provides competitive performance for signal processing. Secondly, it reduces the overall chip area substantially compared to independent DSP and DLA accelerators because of the unified computing arrays and on-chip buffers the majority of the architecture such as on-chip buffers shared across the different applications. Thirdly, the data transfer between signal processing and deep learning can be performed with on-chip buffers without interrupting the GPP using classical mapping optimizations like layer fusion, which benefits the overall system. 


The major contributions of this work can be summarized as follows.
\begin{itemize}
    \item We observe the close correlation of signal processing and deep learning on a broad domain of IoT applications and identify the inefficiency of existing architectures. With this observation, we propose a unified computing architecture, SigDLA, on top of a typical DLA to achieve energy-efficient signal processing and deep learning.
    
    \item SigDLA extends the computing capability of widely used DLAs for signal processing by decoupling the computing array and the on-chip memory with a programmable data shuffling fabric, which converts irregular processing in typical signal processing tasks to tensor processing and enables the deployment of various non-tensor computing tasks. In addition, it has a configurable computing array involved to support variable data width of signal processing and deep learning.

    \item We implement SigDLA on top of NVDLA\cite{NVDLA_website} and achieves an average performance speedup of 4.4$\times$, 1.4$\times$, and 1.52$\times$, and average energy reduction of 4.82$\times$, 3.27$\times$, and 2.15$\times$ compared to an embedded ARM processor with customized DSP instructions, a classical DSP, and an independent DLA-DSP architecture respectively. 
\end{itemize}

%% file: 02_Relatedwork.tex
\section{Related Work \& Motivation}
\subsection{Related Work}

In the ever-evolving era of artificial intelligence (AI), deep learning that dominates existing AI techniques is increasingly applied in IoT devices, and has become a major workload in IoTs processors. The continuously growing importance of deep learning in IoTs stimulates the emergence of many new IoT processors recently. Unlike high-performance processors, IoT processors encounter more severe power, chip area, and performance constraints. They must not only efficiently execute a wide range of deep learning algorithms, but also cater to diverse workloads such as signal processing and data analytics, which presents a new challenge, outpacing the capabilities of classical deep learning accelerators (DLAs). Despite the computing efficiency of typical DLAs, they generally fall short in adaptability for non-AI tasks, highlighting the urgent demand for both efficient and flexible solutions.

An intuitive approach is to reuse general purposed processors and build deep learning frameworks by optimizing deep learning operators. Typical deep learning frameworks such as TinyEngine\cite{mcunet} and Cmix-NN\cite{cmix} have demonstrated significant performance speedup over the direct deep learning processing on MCUs widely used in IoT devices. However, they are usually limited to lightweight models and much less energy-efficient compared to specialized DLAs. A straightforward approach to achieve high energy efficiency is to integrate customized accelerators such as DLAs and DSPs on demand. Hence, for many intelligent IoTs\cite{CPU_DSP_DL} with various sensors, a DLA and a DSP is utilized for deep learning and signal processing respectively. Despite the improved energy efficiency, it takes up more chip area and incurs higher price eventually, which is generally unacceptable for cost-sensitive IoT devices. In addition, when deep learning and signal processing are sequentially utilized for intelligent sensing, they typically need to communicate through the shared memory which poses negative influence on the overall performance and energy efficiency. A relatively more practical solution is to build deep learning engines on top of DSP processors\cite{DSPwithDL1}\cite{DSPwithDL3}\cite{DSPwithDL4}\cite{DSPwithDL5} or extend general purposed processors with customized instructions\cite{fft_shuffle}\cite{shuffling3} optimized for the target computing kernels. These architectures greatly improve the performance of deep learning without compromising the flexibility of the computing engines. Nevertheless, since the baseline architectures i.e. DSP and GPPs are designed for signal processing and generic tasks, the performance for deep learning is generally suboptimal due to the lack of native hardware for deep learning. 

Other than the extension on top of classical processors, coarse-grained reconfigurable arrays (CGRAs)\cite{cascade} \cite{hycube} \cite{CGRAS} that enable rapid runtime reconfiguration for various applications are also explored for computing engines of IoTs. They achieves very good balance between performance and flexibility for a number of different computing kernels. However, deep learning is much more compute-intensive and memory-intensive compared to the signal processing tasks according to the experiments in \ref{sec:motivation}, while CGRAs generally take all the different tasks equally and the controlling overhead is much higher compared to typical DLAs with streamed data flow architecture. Thus, when we take both deep learning workloads and non-deep-learning workloads like signal processing as a whole, a more appropriate approach is to optimize deep learning with higher priority, and the less compute-intensive and memory-intensive workloads with lower priority according to Amdahl's law. In this case, we opt to extend DLA rather than DSP, reusing the DLA architecture to implement some DSP functions without affecting the deep learning workloads.

\subsection{Motivation}\label{sec:motivation}
\begin{figure*}[t]
\centering	
\includegraphics[width=0.65\linewidth]{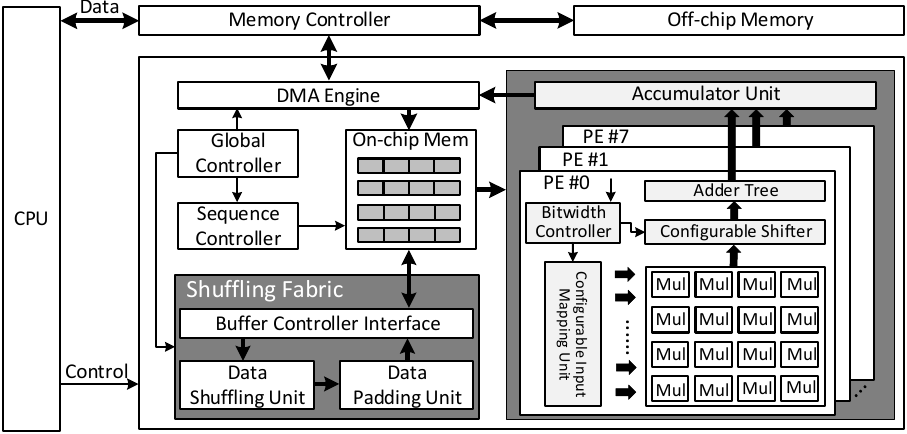}
\caption{SigDLA Architecture Overview.}
\label{fig:overview}
\end{figure*}
As mentioned, signal processing and neural network processing are vital workloads for intelligent sensing in many IoTs. 
Before proceeding with the design of a unified architecture, we investigate the computing requirements of signal processing and neural network processing first. Specifically, we take FFT and FIR as the typical signal processing workloads, and take Tiny-VGGNet\cite{VggTiny} and UltraNet\cite{ultranet} as typical neural network processing tasks. 
We evaluate the computational complexity and parameters of these workloads in Table \ref{tab:workloads comparision}. It can be observed that the computational complexity and the amount of parameters of neural network processing workloads are orders of magnitude higher and larger than those of signal processing workloads. It can be expected that neural network processing will be the performance bottleneck when these workloads are performed at the same time. Hence, the unified architecture should center neural network workloads rather than signal processing. 

\begin{table}[htbp]
  \renewcommand{\arraystretch}{1.2}
  \centering
  \caption{Mult-Adds and Parameters for typical workloads}
    \begin{tabular}{c|c|c|c}
    \hline
    Workloads & Input & Mult-Adds & Parameters \\
    \hline
    radix2-FFT & 1024 complex inputs & 5.12$\times$10$^4$ & 5.12$\times$10$^3$ \\
    \hline
    80-tap FIR & 256 inputs & 2.048$\times$10$^4$ & 80 \\
    \hline
    Tiny-VGGNet & 32 $\times$ 32 $\times$ 3 & 1.69$\times$10$^8$ &  1.15$\times$10$^6$\\
    \hline
    UltraNet & 32 $\times$ 32 $\times$ 3 & 3.83$\times$10$^6$ &  2.07$\times$10$^5$\\
    \hline
    \end{tabular}%
  \label{tab:workloads comparision}%
\end{table}%

The major challenge for signal processing acceleration is the irregular computing pattern, which has been observed in many prior signal processing optimization studies on DSP processors and vector processors \cite{shuffling4} \cite{fft_shuffle} \cite{matrix_shuffling}. To address the problem, software based data shuffling that splits and merges the irregular data sequences for efficient processing on regular computing engines has been proposed. While the software shuffling can induce frequent data transfer between CPUs and the accelerator, we opt to build a hardware shuffling fabric to convert the irregular computing patterns in signal processing to regular ones such that they can be deployed along with the neural network processing on the same regular computing array. In this case, a unified computing architecture can be utilized to sustain both the signal processing and neural network processing efficiently. 


%% file: 03_SigDLA_Architecture.tex
\section{SigDLA Architecture}

In this section, we introduce SigDLA, a unified architecture to support both deep learning and signal processing required by IoT devices with intelligent sensing. As shown in Fig. \ref{fig:overview}, it centers a classical DLA for the regular computing tasks including convolution and GEMMs. On top of the conventional DLA, it incorporates a programmable data reshuffling fabric. This fabric restructures arrays in signal processing algorithms, enabling irregular operations to be efficiently conducted on a regular computing array without affecting deep learning performance.

Specifically, the shuffling fabric is inserted between the data buffer and the computing array to reorganize the shuffled data and convert the processing to regular tensor operations. During the conversion to tensor operations, parts of the tensors need to be padded with fixed values which can be coefficients of signal processing. Therefore, a padding unit is also added to the shuffling structure. The reorganized data will be stored into its original location in the buffer and streamed to the DLA's computing array without breaking the lock-step processing. In this case, the data reorganization is almost transparent to the computing array, which facilitates the reuse of the computing array. While the data shuffling patterns required in typical signal processing algorithms such as FFT and DCT can vary, the shuffling fabric needs to be programmable such that it can be adapted to the different shuffling patterns at runtime. The data shuffling can be controlled with formulated instructions. We extend the traditional DLA tensor operation instructions with our shuffling instructions, allowing both signal processing and deep learning workload to be compiled using the same instruction set. These instructions are streamed to SigDLA via an additional instruction buffer and determine the execution order of the algorithms. The programmable shuffling fabric and control instructions will be detailed in Section \ref{sec:shuffling}. 

As mentioned, another challenge to revisit DLAs for signal processing is the much larger data width used in signal processing, which mainly depends on the precision of the sensors and can vary substantially. The data width of sensors are typically set to be 8-bit, 12-bit, or 16-bit. In contrast, the deep learning models used in IoT devices may be quantized with mixed precision for the sake of less memory overhead and higher computing efficiency. The data width of neural networks generally range from 1-bit to 8-bit. To sustain the computing with distinct data width, we develop a serial computing array based on 4-bit arithmetic operations and it can support higher data width that are multiplies of 4 by reusing the basic 4-bit operations in the computing array. The data width used in the computing array can also be programmed and controlled via our custom tensor instructions. As shown in Fig. \ref{fig:overview}, the implementation of the variable bitwidth computing array includes a bitwidth controller that incorporates bitwidth configuration, as well as configurable input mapping logic and shift logic to generate computation results with variable data widths.

The rest components including DMA engine and sequence controller are basic components of DLAs and could be reused directly. DMA engine is utilized to perform the data transfer between off-chip memory and on-chip memory. The Sequence Controller is responsible for the controlling of the data streamed to the computing array and it is aligned with the execution of the instructions. For the deep learning operations, the sequence controller reads data from the data buffer directly. For the non-deep-learning operations that require data shuffling, it reads data from the data buffer after undergoing the data shuffling fabric.

%% file: 04_Computing_Array.tex
\section{Variable Bitwidth Computing Array}
To accommodate the varying bitwidth requirements of signal processing and deep learning tasks while balancing the computational efficiency and performance of SigDLA, we propose a variable bitwidth computing array. This design draws upon the concept of existing variable bitwidth computing array\cite{BitBlade}\cite{BitFusion} and incorporates shift and addition logic into the 4-bit multiplier, enabling it to support 8-bit or 16-bit multiplication operations. In the following sections, we will delve into the detailed construction of this computing array within SigDLA, and reveal the micro-architecture of the variable bitwidth computing array through the application of corresponding data mapping rules.
\begin{figure*}[t]
    \centering	
    \includegraphics[width=0.85\linewidth]{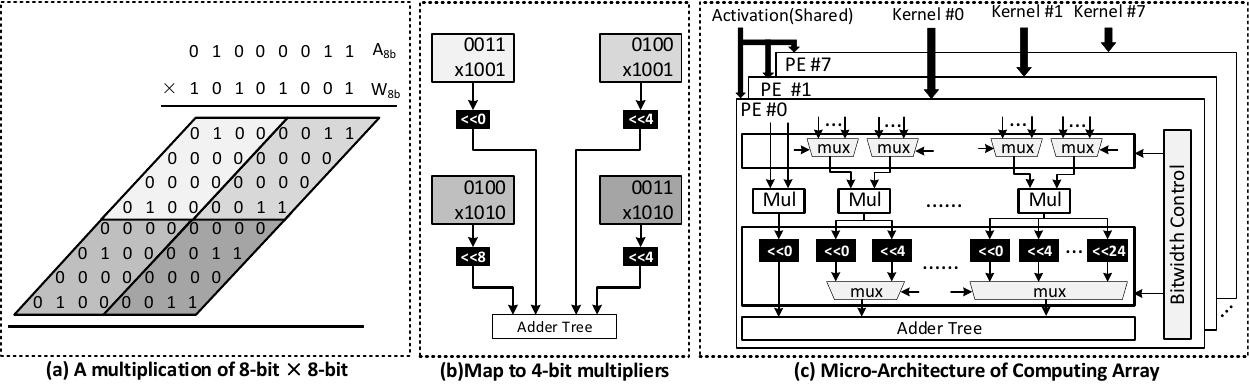}
	\caption{Implementation of Variable Bitwidth Computing Array.}
\label{fig:mac_map}
\end{figure*}
\subsection{Mapping Variable Bitwidth Operations}
To explain how the computing array achieves multiplication under variable bitwidth, the following discussion uses 8-bit multiplication as an example. As mentioned, 8-bit multiplication can be decomposed into 4-bit multiplication. Fig. \ref{fig:mac_map}(a) illustrates the characteristics of multiplying 8-bit operands A$_{\text{8b}}$ and W$_{\text{8b}}$ to produce the final result. The 8-bit multiplication in Fig. \ref{fig:mac_map}(a) is decomposed into four 4-bit multiplications, and the decomposed multiplication is generated using a 4-bit multiplier. The results generated by each 4-bit multiplier need to be shifted before addition. For the 8-bit $\times$ 8-bit case, the shifts of the four multiplications are in order of 0, 4, 4, and 8, as shown in Fig. \ref{fig:mac_map}(b). The same mathematical properties can be recursively applied to 16-bit multiplication. Firstly, the 16-bit multiplication is recursively decomposed into 8-bit multiplication, and then further into 4-bit multiplication. Each level of recursion from 16-bit to 8-bit and from 8-bit to 4-bit requires additional shift-add logic. The next section details the design of a variable-bitwidth computing array that performs variable-bitwidth multiplication and addition using 4-bit multipliers, capable of handling multiplications up to 16-bit.
\subsection{Micro-Architecture}
As shown in Fig. \ref{fig:mac_map}(c), the SigDLA computing array consists of eight precision-scalable PEs, with each PE containing 16 4-bit multipliers. Channel data from pixels or feature maps is mapped into each PE, and all PEs share the same input feature map. The 16 4-bit multipliers inside each PE perform parallel multiplication operations in the input channel direction. The weight for each PE comes from a convolutional kernel, supporting the simultaneous computation of up to eight convolutional kernels. Bitwidth information from the bitwidth controller is sent internally by the global controller. The configurable input mapping logic consists of multiplexers, and the selection signals for the multiplexers are generated by the bit controller's decoding. For different multiplication configurations, the selection signals for the multiplexers have different values. The implementation of the configurable shift logic also depends on the bitwidth configuration information from the bitwidth controller to produce different outputs from the multiplexers. The maximum shift is 24, occurring during a 16-bit $\times$ 16-bit multiplication.

%% file: 05_Data_Shuffling.tex
\section{Programmable Data Shuffling}\label{sec:shuffling}
This section analyzes the methods for implementing data shuffling in the DLA. The DLA computing array can efficiently handle matrix operations. By mapping signal processing algorithms to convolutional layers, the DLA acquires signal processing capabilities. We have observed that the data sequences in the convolutional layers of the DLA exhibit certain regularities, and signal processing algorithms can be mapped to convolutional layers through data shuffling. However, the required data shuffling rules vary for different signal processing algorithms. Establishing a universal data shuffling logic is crucial for enabling the DLA to support arbitrary signal processing algorithms. In the following, we will delve deeper into the key technologies for implementing signal processing in the DLA.

\subsection{Mapping Signal Processing Operations to Convolution}
The signal processing algorithms, such as FFT, DCT, FIR, and DWT, are not regular matrix operation formats, but many signal processing algorithms can be transformed into matrix operations after processing\cite{shuffling3}\cite{fft_shuffle}\cite{fft_cvt_matrix}, and have some similarities\cite{DSPwithDL1}\cite{FFT_CONV_Sim} with convolution operations in CNN.
\begin{figure}[htb]
    \centering	
    \includegraphics[width=0.9\linewidth]{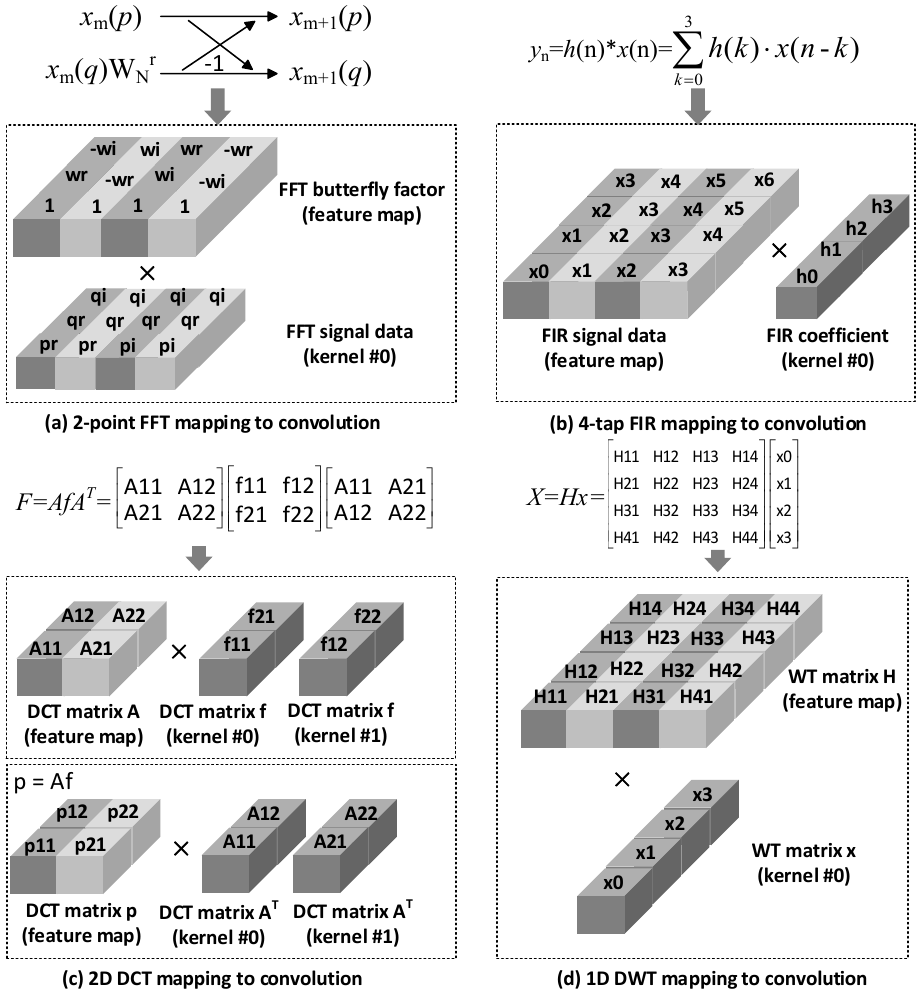}
	\caption{Mapping different signal processing algorithms to convolution.}
\label{fig:mac_sig}
\end{figure}
It can be seen from Fig. \ref{fig:mac_sig}(a) that in the butterfly operation of the 2-point FFT, the butterfly factor is mapped to the feature map part of the convolution layer, and the signal data is mapped to the convolution kernels, where $wr$ and $wi$ are the real and imaginary parts of the butterfly factor, $qr$ and $qi$ are the real and imaginary parts of $x_m(q)$, $pr$ and $pi$ are the real and imaginary parts of $x_m (p)$. Fig. \ref{fig:mac_sig}(b) shows the schematic diagram of FIR mapping to convolution layer. The input part $x$ of FIR is mapped to the feature map and $h$ is mapped to the convolution kernel. Fig. \ref{fig:mac_sig}(c) and (d) show the DCT algorithm and the DWT algorithm. Their regular matrix operations can be efficiently mapped to convolution layer.

\subsection{Micro-Architecture}
\begin{figure*}[t]
\centering	\includegraphics[width=0.9\linewidth]{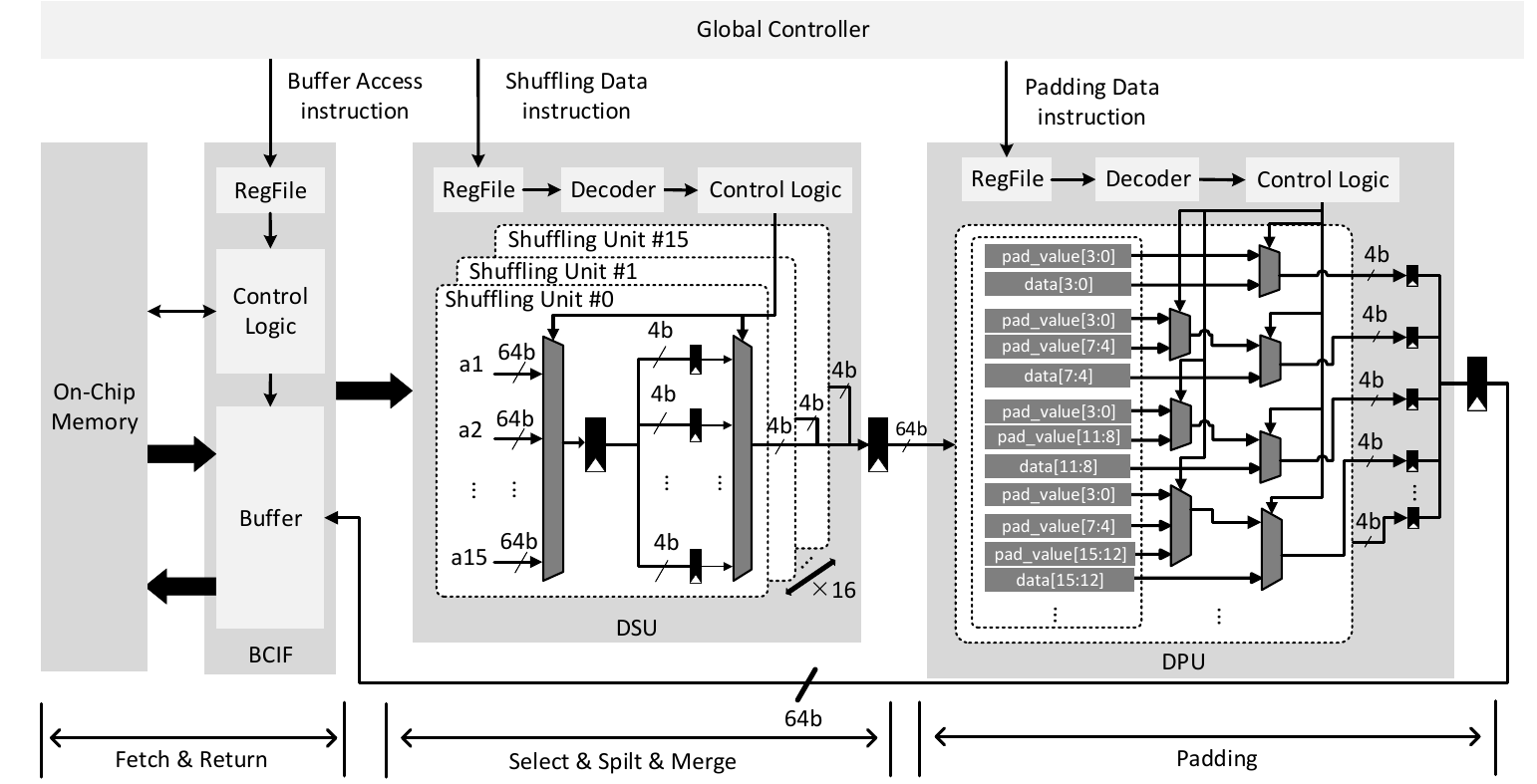}
	\caption{The Micro-Architecture of Shuffling Fabric}
\label{fig:prog_shuffling}
\end{figure*}
As previously analyzed, the core challenge in implementing signal processing on the DLA lies in the artful transformation of irregular arithmetic operations into matrix operations within the convolutional layers of a CNN. This transformation is necessary because the original data formats used in signal processing algorithms are often complex and irregular, making it challenging to directly adapt them to the operational mode of the DLA. Therefore, an effective shuffling mechanism must be designed to convert the original data into a matrix format that is compatible with CNN processing, enabling efficient and accurate signal processing. This section provides a thorough exposition of the micro-architecture of the shuffling fabric. Fig. \ref{fig:prog_shuffling} shows the micro-architecture of the shuffling fabric, where the shuffling fabric comprises a Buffer Controller Interface (BCIF), a Data Shuffling Unit (DSU), and a Data Padding Unit (DPU). In the following sections, we will delve into the functionality and operation of each of these modules in detail.
\subsubsection{Buffer Controller Interface}
The BCIF includes read control logic and write control logic, as well as a register file containing configuration information. The register file is used to store instructions that are sent to the global controller through the top-level port of the module by the Host Processor. The global controller then distributes the instructions to the register file within the BCIF based on address allocation. The read control logic generates corresponding read addresses and read sequence lengths based on the configuration of the register file. The write control logic writes the post-processed data back to the original address after the DPU completes its work. The write control logic needs to specify the data type being written back to the on-chip memory. The BCIF incorporates a data buffer unit to store a certain amount of data for use by the DSU. Typically, pre-fetched data is divided into two parts, such as feature map data and weights in deep learning, or preprocessed signals and weights in signal processing. These two parts of data are stored in separate continuous bank units following different starting addresses. 

\subsubsection{Data Shuffling Unit}
The DSU retrieves data from the BCIF data buffer and shuffles it accordingly. The DSU includes a register file that stores the configurations for the data reshuffling process. The shuffling logic is implemented through a shuffling array, which comprises 16 shuffling units, each with identical functionality. The shuffling unit selects one data from 16 64-bit input data through the first multiplexer. This selected data is then separated into 4-bit units and stored sequentially in 16 registers. Subsequently, the second multiplexer selects one data from these 16 registers and places it in a specific 4-bit position of a new 64-bit register. Altogether, the 16 shuffling units can process sixteen 64-bit data in parallel, with each shuffling unit outputting a 4-bit data. By connecting the outputs of all 16 shuffling units, a new 64-bit data is obtained.

\subsubsection{Data Padding Unit}
The DPU is primarily responsible for padding operations on shuffled data. Some signal processing algorithms, such as the butterfly operation in FFT, require specific positions to be filled with a fixed value of ``1" after being converted into matrix operations. The DPU is capable of padding specific constants within the matrix after signal processing algorithms like FFT are converted into matrix operations. For a 64-bit data, when the bitwidth is 4-bit, 8-bit, and 16-bit, the number of valid padding position for the 64-bit data is 16, 8, and 4, respectively. The effective bitwidth of the padding values is 16-bit, 8-bit, and 4-bit, in order. The padding process is influenced by the bitwidth. After receiving data from the DFU, the DPU generates processed data based on bitwidth configuration information, the position of padding, and the padding value information stored in the register file.

\subsection{Shuffling Instructions}
\begin{figure}[htbp]
    \centering	
    \includegraphics[width=0.95\linewidth]{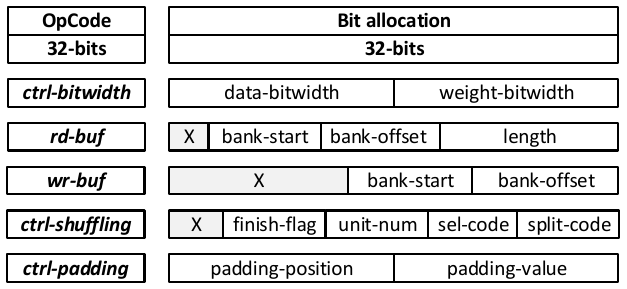}
	\caption{Shuffling Instruction.}
\label{fig:shuffling_inst}
\end{figure}

This section explains the implementation of the instruction corresponding to the programmable data shuffling hardware. These instructions provide a software-level abstraction, allowing programs written by the CPU to conveniently utilize data shuffling techniques, effectively implementing various signal processing algorithms on the SigDLA. As shown in Fig. \ref{fig:shuffling_inst}, the functions of the instructions can be divided into the following sections.

Managing memory access for BCIF. The \textit{rd-buf/wr-buf} instructions control the reading and writing of on-chip memory. The \textit{rd-buf} instruction occurs before data shuffling, used to read the required amount of data into the BCIF. The \textit{wr-buf} instruction occurs after shuffling, writing the data back to the specified location in on-chip memory. The bank-start and bank-offset generate address information, while length determines the number of read sequences. 

Control the bitwidth configuration of the SigDLA. The \textit{ctrl-bitwidth} instruction is used to specify the bitwidth of operands to ensure correct data processing and computation. Modules that utilize bitwidth include the variable-bitwidth computing array of the SigDLA and the data padding unit.

Configure the DSU to generate specific shuffling rules. The \textit{ctrl-shuffling} instruction controls the rules of data shuffling. It selects one of the sixteen units in the DSU using unit-num and controls the unit's behavior using sel-code and split-code. The finish-flag is used to determine if all the units currently required for the task have been fully configured. 

Control the DPU to padding data. The \textit{ctrl-padding} instruction controls the rules of data padding. The padding configuration, including padding-position and padding-value, written through \textit{ctrl-padding}, is used to select the location and value for data padding.

\begin{figure}[htbp]
    \centering	
    \includegraphics[width=0.85\linewidth]{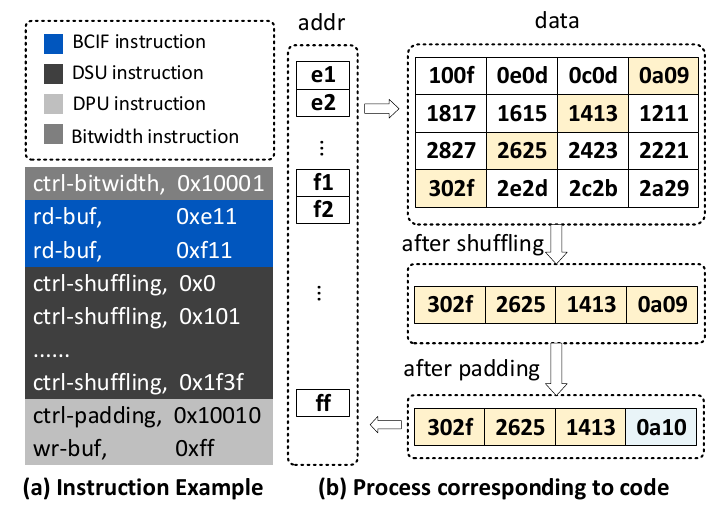}
	\caption{An example of data shuffling using instructions.}
 \vspace{-1em}
\label{fig:shuffling_example}
\end{figure}

Fig. \ref{fig:shuffling_example} shows a case study. Four data items are retrieved from the on-chip memory using the \textit{rd-buf} instruction. Based on \textit{ctrl-shuffling}, four 16-bit data segments are extracted from the four 64-bit data items and recombined into a new data item. Subsequently, the lowest 8 bits are padded using the \textit{ctrl-padding} instruction, and finally, the new data item is written back to the on-chip memory through the \textit{wr-buf} instruction.

%% file: 06_Evaluation.tex
\section{Evaluation}
\subsection{Experiment Setup}
We use Verilog to implement SigDLA on the basis of small-NVDLA,  and we have developed a cycle accurate simulator for SigDLA, providing a high-precision simulation environment for algorithm performance evaluation. We use Synopsys Design Compiler to synthesize SigDLA at the UMC 55nm technology node. Design Compiler provides the chip area, frequency, and power consumption. UltraNet\cite{ultranet}, Tiny-VGGNet\cite{VggTiny}, ResNet20\cite{ResNet50}, FFT, 2D-DCT, and FIR were selected as benchmarks to evaluate the improvement in performance of SigDLA with variable bitwidth. When comparing different hardware platforms, we selected ARM Cortex-M4 embedded processor and TMS320F28x\cite{tms320f28x} digital signal processor, and used FFT and FIR as benchmarks to evaluate the performance and energy reduction of SigDLA in signal processing algorithms. Under intelligent IoT, we chose deep learning algorithms \cite{DSP_DLA_task} for signal analysis as the benchmark evaluation and compared SigDLA with independent DSP-DLA architectures. During the performance evaluation, all the hardware involved in the comparison adopted a clock frequency of 100MHZ. The performance and power consumption data of ARM Cortex M4 were obtained based on the MAX78000 development kit\cite{MAX78000}, while the performance and power consumption data of TMS320F28x were obtained based on the TMS320F28335 development kit.

\subsection{System Specifications}
As shown in Table \ref{tab:Performance comparison}, using UMC 55nm technology for synthesis, SigDLA has a chip area of 5.21mm$^2$, a leakage power consumption of 2.02mW at a working voltage of 1.2V, and a total power consumption of 302.5mW. The total size of on-chip memory is 144KB, of which 16KB is dedicated to signal processing algorithms. Compared to small-NVDLA, the chip area of SigDLA has increased by 17\%, and the total power consumption has increased by 9.4\%. SigDLA supports signal processing algorithms such as FFT, FIR, and DCT, which are not supported by small-NVDLA. SigDLA supports 4-bit, 8-bit and 16-bit data types, while small-NVDLA only supports 8-bit.
\vspace{-0.5em}
\begin{table}[htbp]
  \renewcommand{\arraystretch}{1.2}
  \centering
  \caption{Hardware overhead comparison between small-NVDLA and SigDLA}
    \begin{tabular}{c|c|c}
    \hline
         & small-NVDLA & SigDLA \\
    \hline
    Technology & 55nm & 55nm \\
    \hline
    Core Area(mm$^2$) &   4.45   & 5.21 \\
    \hline
    Frequency(MHz) &   100   &  100\\
    \hline
    On-chip memory  & 128KB & 128KB + 16KB \\
    \hline
    Voltage(V)  & 1.2V     & 1.2V \\
    \hline
    Total Power(mW) &   276.4   &  302.5\\
    \hline
    Leakage(mW) &   1.72   &  2.02\\
    \hline
    Data Types(Bit) & 8-bit & 4-bit, 8-bit, 16-bit \\
    \hline
    Algorithm Support & DNN  & DNN, DSP \\ 
    \hline
    \end{tabular}%
  \label{tab:Performance comparison}%
\end{table}%

\subsection{Performance and Energy Comparison}
\subsubsection{Variable-bitwidth Performance Comparison}
\begin{figure}[htbp]
\centering	\includegraphics[width=0.95\linewidth]{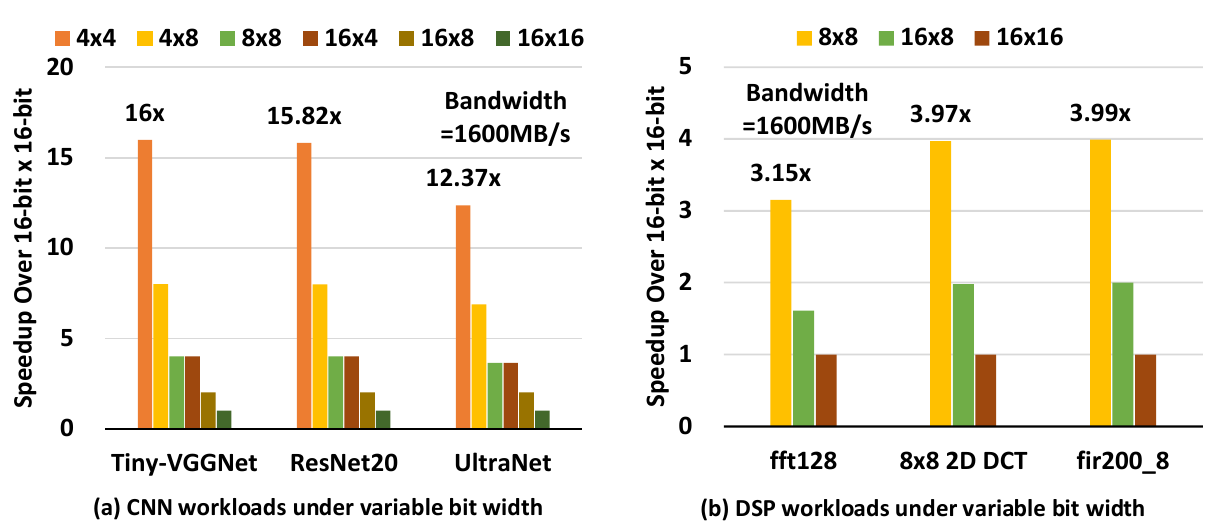}
	\caption{Variable-bitwidth speedup on CNN and DSP workloads.}
 \vspace{-1em}
\label{fig:result_bitwidth}
\end{figure}

Based on the SigDLA simulator, we tested the variable bitwidth benchmark in detail in the 100MHz simulation environment. For the CNN benchmark, when the input is 32 $\times$ 32 $\times$ 3, the experimental results show that SigDLA shows the shortest inference time under 4-bit $\times$ 4-bit. As shown in Fig. \ref{fig:result_bitwidth}(a), at a typical frequency of 100MHz, the bandwidth of off-chip memory is set to 1600MB/s\cite{dram_speed}. TinyVGG-Net, ResNet20 and UltraNet achieve 16$\times$, 15.82$\times$ and 12.37$\times$ speedup at 4-bit $\times$ 4-bit compared with 16-bit $\times$ 16-bit. For the DSP benchmark, as shown in Fig. \ref{fig:result_bitwidth}(b), the benchmark of DSP is less affected by bandwidth, because its parameter quantity is far less than that of deep learning algorithm. The 128 point complex FFT, 2D-DCT and 200 point 8-taps FIR achieve 3.15$\times$, 3.97$\times$ and 3.99$\times$ speedup than 16-bit $\times$ 16-bit at 8-bit $\times$ 8-bit. The speedup of FFT is significantly lower than that of DCT and FIR, mainly because more shuffling operations are required for converting FFT to convolution operations and the computational complexity of FFT is higher.
\subsubsection{Signal Processing Algorithm Comparison}
\begin{figure}[htbp]
\centering	\includegraphics[width=0.9\linewidth]{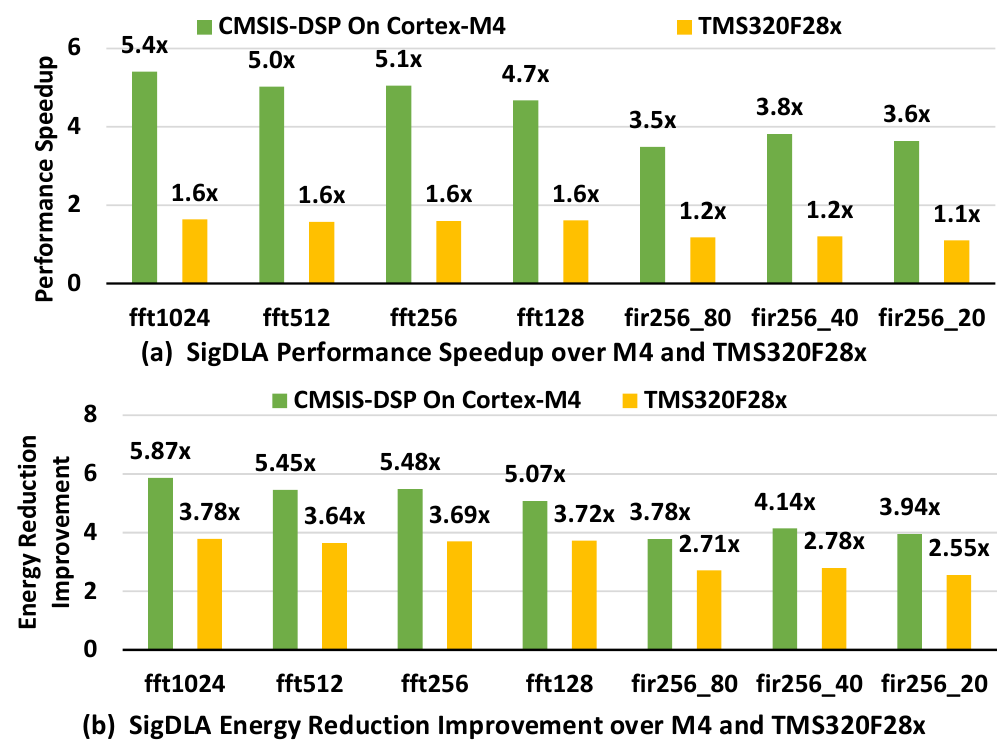}
	\caption{Performance and Energy Reduction of signal processing algorithms.}
\label{fig:result_signal}
\end{figure}

To more accurately evaluate the performance and power consumption of SigDLA in signal processing algorithms, we conducted a thorough comparison between it and two processors. Among them, ARM Cortex-M4 utilizes the CMSIS-DSP library to run signal processing algorithms. As shown in Fig. \ref{fig:result_signal}, we selected the FFT and FIR algorithms for testing in signal processing. For the FFT algorithm test, we employed 16-bit complex inputs and evaluated performance at 1024 points, 512 points, 256 points, and 128 points. Regarding the FIR algorithm, we tested the performance of a 256-point sampled signal with filter taps of 20, 40 and 80. After comparison, we found that SigDLA outperformed both TMS320F28x and ARM Cortex-M4 in terms of FFT and FIR algorithms. Specifically, SigDLA achieved an average performance speedup of 1.4$\times$ and 3.27$\times$ energy reduction compared to TMS320F28x, while compared to ARM Cortex-M4, SigDLA achieved a performance speedup of 4.4$\times$ and an energy reduction of 4.82$\times$.

\subsubsection{CNN-Based Signal Processing Algorithm Comparison}

\begin{figure}[htbp]
\centering	
    \includegraphics[width=0.8\columnwidth]{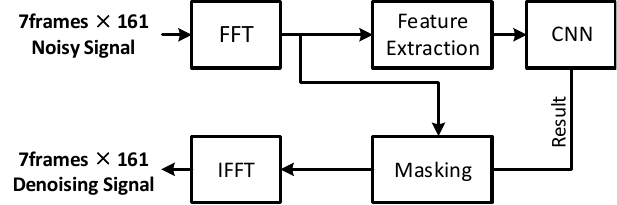}
	\caption{CNN-based speech enhancement algorithm.}
\label{fig:dsp_dla_task}
\end{figure}

The core objective of our design is to enhance the energy efficiency of intelligent IoT devices that concurrently use digital signal processing and deep learning analysis. Therefore, we utilize the CNN-Based Signal Processing Algorithm to test our design. As shown in Fig. \ref{fig:dsp_dla_task}, the input speech signal is first processed by FFT algorithm and converted to frequency domain. Then, the feature of the processed signal is extracted and input into the subsequent CNN model. CNN model generates a mask that can effectively shield the noise in the speech signal and significantly improve the speech intelligibility. Then, the denoised signal is converted back to the time domain to present a better sound effect.

\begin{figure}[htbp]
\centering	
    \includegraphics[width=0.9\linewidth]{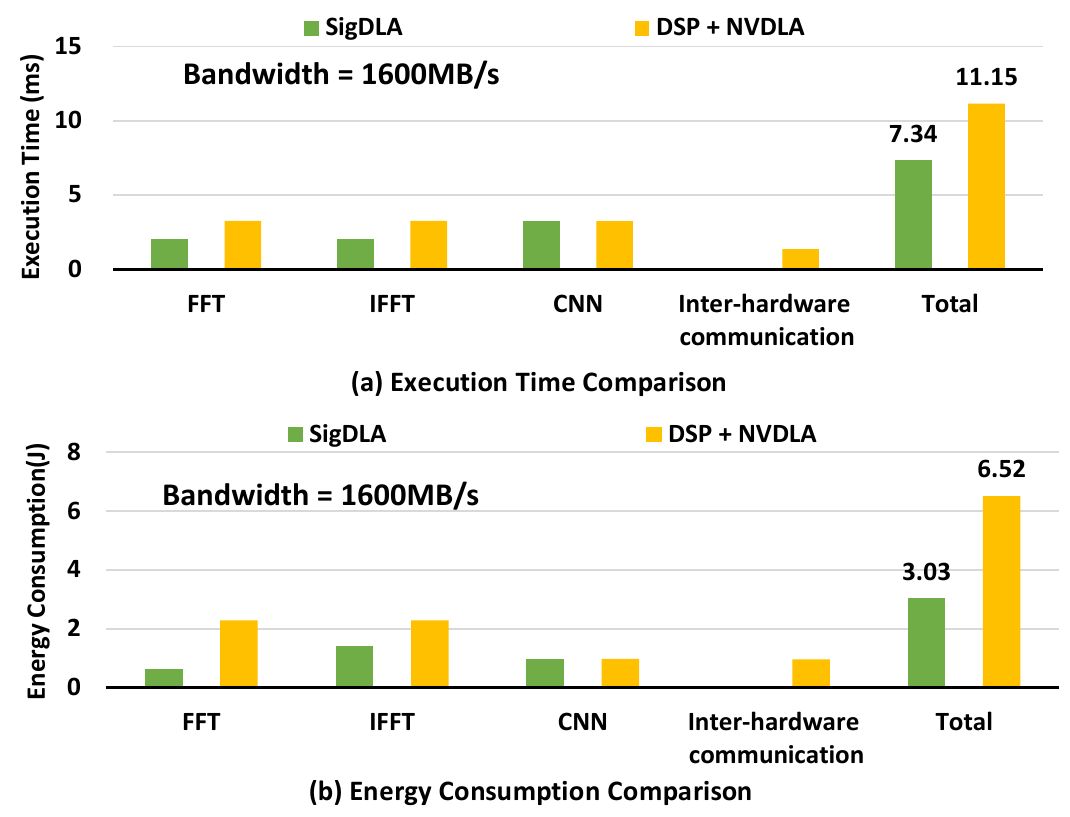}
	\caption{Performance and Energy Consumption of CNN-Based Signal Processing Algorithm.}
\label{fig:result_dsp_dla}
\end{figure}

To evaluate the performance and power consumption of different architectures in handling this task, we chose SigDLA and an independent DSP-DLA architecture for comparison. The independent DSP-DLA architecture combines the TMS320F28x processor and small-NVDLA. In the processing of FFT algorithm, we use 8-bit data type. For SigDLA, we use 8-bit pixel format and 4-bit weight format, while small NVDLA uses the 8-bit $\times$ 8-bit data type. The independent DSP-DLA architecture requires the FFT results calculated by the TMS320F28x processor to be written into off-chip memory during processing, which is then read by small-NVDLA. This process involves data transmission and storage. In contrast, SigDLA is continuous in the switching process between FFT and CNN, without writing data to off-chip memory and then reading, thus reducing the overhead of data transmission. As shown in Fig. \ref{fig:result_dsp_dla}, SigDLA achieves 1.52$\times$ speedup and 2.15$\times$ energy reduction than the independent DSP-DLA architecture in the coexistence of signal processing analysis and deep learning. This remarkable acceleration effect is mainly due to the fact that SigDLA does not need to communicate between hardware and its efficient signal processing speed.

%% file: 07_Conclusion.tex
\section{Conclusion}
In this paper, we present a unified computing architecture SigDLA based on a typical DLA to achieve efficient signal processing and deep learning that are typically required in many intelligent sensing scenarios. While signal processing like FFT usually involves many irregular data shuffling and computing and cannot be directly applied on a typical DLA targeting only regular operations like convolution and GEMMs, we propose an online data shuffling fabric to convert the irregular operations within signal processing to regular tensor operations such that typical signal processing tasks can also be implemented on the same computing array of DLAs. Moreover, we also leverage a flexible computing array with variable bit width such as 4-bit, 8-bit, and 16-bit to suit the diverse data width requirements of both deep learning and signal processing. According to our experiments on a set of signal processing and deep learning tasks, SigDLA achieves an average performance speedup of 4.4$\times$, 1.4$\times$, and 1.52$\times$, and an average energy reduction of 4.82$\times$, 3.27$\times$, and 2.15$\times$ compared to an embedded ARM processor with customized DSP instructions, DSP processor, and independent DSP-DLA architecture, while it takes only 17\% more chip area than the original DLA.